\documentclass[aps,10pt,showkeys,floatfix,superscriptaddress]{revtex4}

\usepackage{times}
\usepackage{amsfonts}
\usepackage{amsmath}
\usepackage{amssymb}
\usepackage{graphicx}
\usepackage{epsfig}

\newcommand{\Rea}{\mathop{\mathrm{Re}}\nolimits}
\newcommand{\Ima}{\mathop{\mathrm{Im}}\nolimits}

\begin{document}

\title{Scattering and Transformation of Waves on Heavy Particles in Magnetized Plasma}
\author{Hrachya B. Nersisyan}
\email{hrachya@irphe.am}
\affiliation{Institute of Radiophysics and Electronics, 0203 Ashtarak, Armenia}
\affiliation{Centre of Strong Fields Physics, Yerevan State University, Alex Manoogian
str. 1, 0025 Yerevan, Armenia}
\date{\today }

\begin{abstract}
The scattering and transformation of the waves propagating in magnetized
plasma on a heavy stationary charged particle located at a plane
plasma-vacuum boundary is considered. The scattering (transformation) occurs
due to the nonlinear coupling of the incident wave with the polarization
(shielding) cloud surrounding the particle. It is shown that the problem is
reduced to the determination of the nonlinear (three index) dielectric
tensor of magnetized plasma. The angular distribution and the cross section
for scattering (transformation) of high-frequency ordinary and extraordinary
waves and low-frequency upper-hybrid, low-hybrid, and magnetosonic waves are
investigated within a cold plasma (hydrodynamic) model.
\end{abstract}

\maketitle

\section{Introduction}
\label{sec:1}

It is well known that in a medium with a certain fluctuation level, the
propagation of electromagnetic waves can lead to radiation of waves with new
frequencies and wave numbers, i.e. scattered waves and also a new type of
wave: transformed waves. The investigations of electromagnetic waves
scattering and transformation processes are very important for studying such
problems as plasma diagnostics, wave transformation mechanisms in plasma,
definition of dispersion properties of plasma wave processes etc. In
addition the study of the electromagnetic wave scattering spectra (both in
laser and microwave wave ranges) is an efficient method of plasma
diagnostics in laboratory fusion research devices as well as in the near and
outer space.

Electromagnetic wave scattering is caused by thermal fluctuations of plasma
density and other plasma parameters such as current density, electric and
magnetic fields, etc. Spectra of scattered waves provide information on the
density and temperature distributions in the plasma. A peculiarity of
electromagnetic wave scattering in plasmas is coherent scattering by
collective plasma excitations-combination scattering that occurs along with
Thompson incoherent scattering by individual plasma particles. Wave
scattering by collective plasma fluctuations, in particular, makes it
possible to find relative concentrations of charged particles and
temperatures of individual plasma components. The phenomenon of
electromagnetic wave combination scattering by collective plasma excitations
has been considered for the first time by Akhiezer et al. (see, e.g., \cite{1}).
Subsequently a theory of electromagnetic wave scattering in plasmas has been
developed \cite{1,2,3,4,5,6,7}. The detailed theory of scattering and transformation of
waves in magnetized plasma has been worked out in \cite{8} (see also \cite{9,10} where
useful reviews of the electromagnetic wave scattering problem have been
presented). The theory has been further developed and extended in the papers
\cite{11,12,13,14,15} (see also references therein), in particular, in the case of
strongly magnetized turbulent plasma \cite{14}. The scattering and transformation
of high-frequency waves in dusty plasmas due to electron density
inhomogeneities have been investigated in \cite{15}. The scattering and
transformation cross sections for two cases (the electrons in the shielding
clouds around the charged dust particles and induced electron density
fluctuations discreteness) have been calculated and it has been shown that
both can be enhanced with respect to scattering from thermal fluctuations.

Another important mechanism for the wave scattering and transformation could
be provided by the nonlinear interaction of the incident wave with the
non-thermal density fluctuations (wakefield excitations) generated by the
charged test particles moving in plasma. In particular, if such particle is
at rest and heavy and does not oscillate in the electromagnetic field of the
incident wave the scattering occurs due to the nonlinear oscillation of the
polarization (shielding) cloud surrounding the particle. It is clear that
the cross section of this process essentially differs from the standard
Thomson cross section when the wavelength of the incident wave is comparable
or smaller the size of the polarization cloud which is typically given by
the Debye screening length. Moreover, in a nonlinear regime the polarization
cloud surrounding the particle may be affected by an external strong
magnetic field which introduces a strong anisotropy in the screening
properties of the plasma and as a result in angular distribution of the
scattered (transformed) waves.

In this paper we investigate this process in detail assuming that the heavy
test particle is located at the plane boundary of plasma-vacuum interface.
The plasma is assumed to be strongly magnetized so that the cyclotron
frequency of the electrons is comparable or even larger than the plasma
frequency. The case in which the incident wave propagates across the
external magnetic field is considered. Within cold plasma model general
expressions are obtained for the angular distribution and the cross section
for the scattered and transformed waves. The explicit calculations are done
for specific high-frequency ordinary and extraordinary waves as well as for
low-frequency upper-hybrid, lower-hybrid, and magnetosonic waves.

\section{Theoretical Model}
\label{sec:2}

The main problem in calculating the quantitative characteristics of electromagnetic wave scattering and transformation
in plasmas is to find the current produced by the nonlinear interaction of the incident wave with the fluctuations of
plasma parameters caused by the test particle. This current determines the scattered and transformed wave fields.

We consider an incident wave $\mathbf{E}^{(0)} =\boldsymbol{\mathcal{E}}_{0} e^{i\mathbf{k}_{0} \cdot \mathbf{r}-i\omega
_{0} t} +\text{c.c.}$ (where $\boldsymbol{\mathcal{E}}_{0}$ is the complex amplitude) which propagates in magnetized
homogeneous plasma and a heavy particle with charge $Ze$ ($-e$ is the charge of an electron) at rest. The amplitude of
the magnetic field of the incident wave is determined by the Maxwell's equation and has the form $\boldsymbol{\mathcal{B}}%
_{0} =(c/\omega _{0} ) \lbrack \mathbf{k}_{0} \times \boldsymbol{\mathcal{E}}_{0} \rbrack$. In the linear approximation the
incident wave and the electric field $\delta \mathbf{E}(\mathbf{r})$ produced by the test particle are independent, and
the Fourier transformed total electric field in plasma is
\begin{equation}
\mathbf{E}^{(1)} (\mathbf{k},\omega )=\boldsymbol{\mathcal{E}}_{0} \delta \left( \mathbf{k}-\mathbf{%
k}_{0} \right) \delta \left( \omega -\omega _{0} \right)
+\boldsymbol{\mathcal{E}}_{0}^{*} \delta \left( \mathbf{k}+\mathbf{k}_{0} \right) \delta \left(
\omega +\omega _{0} \right) +\delta \mathbf{E}(\mathbf{k})\delta (\omega ) .
\label{eq:1}
\end{equation}

The amplitude and the frequency of the incident wave, for given values of the wave vector, are determined from the equations
$M_{ij} (\mathbf{k}_{0} ,\omega _{0} ) \mathcal{E}_{0j} =0$ and $\det \left| M_{ij} (\mathbf{k}_{0} ,\omega _{0} )\right| =0$,
respectively, where
\begin{equation}
M_{ij} (\mathbf{k},\omega )=\delta _{ij} -\frac{\omega ^{2} }{k^{2} c^{2} }
\varepsilon _{ij} (\mathbf{k},\omega )-\frac{k_{i} k_{j} }{k^{2} }
\label{eq:2}
\end{equation}%
and $\delta _{ij}$ are the Maxwellian and unit tensors, respectively, and $\varepsilon _{ij} (\mathbf{k},\omega )$ is the
dielectric tensor of the magnetized plasma.

The electric field of the stationary heavy particle is expressed by the equation \cite{16,17}
\begin{equation}
\delta \mathbf{E}(\mathbf{k})=-\frac{4\pi iZe\mathbf{k}}{(2\pi )^{3} k^{2}
\varepsilon (\mathbf{k},0)} .
\label{eq:3}
\end{equation}%
Here $\varepsilon (\mathbf{k},\omega )=k_{i} k_{j} \varepsilon _{ij} (\mathbf{k} ,\omega )/k^{2}$ is the longitudinal
dielectric function of the plasma. We assume that the particle is heavy and does not oscillate in the field of the incident
wave. Thus the scattering originates from oscillations of the polarization cloud surrounding the particle.

To find the electromagnetic field of the scattered (transformed) wave, we consider the second order approximation in Maxwell's
equations. As a result, for the electric field $E_{j}^{(2)} (\mathbf{k},\omega )$ in the second order approximation we obtain the
equation
\begin{equation}
M_{ij} (\mathbf{k},\omega )E_{j}^{(2)} (\mathbf{k},\omega )=\frac{4\pi
i\omega }{k^{2} c^{2} } J_{i} (\mathbf{k},\omega ) ,
\label{eq:4}
\end{equation}%
where
\begin{equation}
J_{i} (\mathbf{k},\omega )=\frac{\omega }{4\pi i} \int d\mathbf{%
k^{\prime} } \int d\omega ^{\prime} \varepsilon _{ijk} (\mathbf{k}%
,\omega ;\mathbf{k^{\prime} },\omega ^{\prime} )
E_{j}^{(1)} (\mathbf{k^{\prime \prime} },\omega ^{\prime \prime}
)E_{k}^{(1)} (\mathbf{k^{\prime} },\omega ^{\prime} )
\label{eq:5}
\end{equation}
is the nonlinear current associated with the nonlinear, three-index dielectric tensor $\varepsilon _{ijk} (\mathbf{k},\omega%
;\mathbf{k^{\prime} },\omega ^{\prime})$ of the magnetized plasma, $\mathbf{k^{\prime \prime} }=\mathbf{k}-\mathbf{k^{\prime} }$
and $\omega ^{\prime \prime} =\omega -\omega ^{\prime}$.

The scattered waves originate from nonlinear coupling of the incident wave with the electric field of the test particle. The
scattering current corresponding to such coupling is easily obtained from Equations \eqref{eq:1} and \eqref{eq:5} if in the
expression obtained
for the current we neglect the terms proportional to $\mathcal{E}_{0j} \mathcal{E}_{0k}$ and $\delta E_{j} (\mathbf{k^{\prime %
\prime} })\delta E_{k} (\mathbf{k^{\prime} })$, which determine the second harmonic generation and the second order electric
field of the test particle, respectively. The total scattering current is thus determined by the equation
\begin{eqnarray}
J_{i}^{(s)} (\mathbf{k},\omega )=\frac{\omega }{4\pi i} \left[ S_{ijl}
\left( \mathbf{k},\omega _{0} ;\mathbf{k}_{0} ,\omega _{0} \right) \delta
E_{j} \left( \mathbf{k}-\mathbf{k}_{0} \right) \mathcal{E}_{0l} \delta \left( \omega -\omega _{0} \right) \right. \label{eq:6} \\
\left. +S_{ijl} \left( \mathbf{k},-\omega _{0} ;-\mathbf{k}_{0} ,-\omega _{0} \right)
\delta E_{j} \left( \mathbf{k}+\mathbf{k}_{0} \right)
\mathcal{E}_{0l}^{*} \delta \left( \omega +\omega _{0} \right) \right] ,  \nonumber
\end{eqnarray}
where the tensor $S_{ijl}$ characterizes the nonlinear properties of the medium \cite{18}:
\begin{equation}
S_{ijl} \left( \mathbf{k},\omega ;\mathbf{k^{\prime} },\omega ^{\prime}
\right) =\varepsilon _{ijl} \left( \mathbf{k},\omega ;\mathbf{k^{\prime} }%
,\omega ^{\prime} \right) +\varepsilon _{ilj} \left( \mathbf{k},\omega ;%
\mathbf{k^{\prime \prime} },\omega ^{\prime \prime} \right) .
\label{eq:7}
\end{equation}%

The electric field $E^{\prime} _{i} (\mathbf{k},\omega )$ of the scattered wave is obtained from Maxwell's equation \eqref{eq:4},
in which the current $\mathbf{J}(\mathbf{k},\omega )$ is replaced by the scattering current $\mathbf{J}^{(s)} (\mathbf{k},\omega )$.
Thus
\begin{equation}
E^{\prime} _{i} (\mathbf{k},\omega )=\frac{4\pi i\omega }{k^{2} c^{2} }
T_{ij} (\mathbf{k},\omega )J_{j}^{(s)} (\mathbf{k},\omega ) .
\label{eq:8}
\end{equation}%
Here $T_{li} (\mathbf{k},\omega )$ is the tensor inverse to the Maxwellian tensor, $T_{li} (\mathbf{k},\omega )M_{ij} (\mathbf{k},%
\omega )=\delta _{lj}$.

Since the intensity $W_{s}$ of the scattered radiation is equal to (with the minus sign) the work performed by the source of the
scattered radiation per unit time, neglecting damping of the scattered wave, we obtain
\begin{equation}
W_{s} =-\frac{2iZ^{2} e^{2} \omega _{0}^{3} \left| \boldsymbol{\mathcal{E}}_{0} \right|
^{2} }{(2\pi )^{2} c^{2} } \int \frac{d\mathbf{k}}{k^{2} } \frac{%
A_{i} (\mathbf{k})A_{j}^{*} (\mathbf{k})}{(\mathbf{k}-\mathbf{k}_{0} )^{4}
\left| \varepsilon \left( \mathbf{k}-\mathbf{k}_{0} ,0\right) \right| ^{2} }
\Ima \left[ T_{ji} (\mathbf{k},\omega _{0} )-T_{ij}^{*} (\mathbf{k},\omega _{0} )\right] ,
\label{eq:9}
\end{equation}
where $A_{i} (\mathbf{k})=S_{isl} (\mathbf{k},\omega _{0} ;\mathbf{k}_{0} ,\omega _{0} )(k_{s} -k_{0s} )e_{l}$, $\mathbf{e}=
\boldsymbol{\mathcal{E}}_{0} /|\boldsymbol{\mathcal{E}}_{0} |$ being the complex unit vector along the direction of the
polarization of the incident wave. As one would expect, it is seen from Eq.~\eqref{eq:9} that the scattering (transformation) on a
stationary charge occurs with no frequency change ($\omega ^{\prime} =\omega _{0}$).

The total scattering cross section $\sigma$ is the ratio of the intensity $W_{s} $ of the scattered radiation to the energy flux
$\mathbf{S}=(c|\boldsymbol{\mathcal{E}}_{0} |^{2} /2\pi )\mathbf{S}_{0}$ in the incident wave, where
\begin{equation}
\mathbf{S}_{0} =\frac{\mathbf{v}_{g} }{2\omega _{0} c} \frac{\partial }{%
\partial \omega _{0} } \left[ \omega _{0}^{2} e_{i} e_{j}^{*} \varepsilon
_{ij}^{\text{(H)} } \left( \mathbf{k}_{0} ,\omega _{0} \right) \right] .
\label{eq:10}
\end{equation}%
Here $\varepsilon _{ij}^{\text{(H)} } (\mathbf{k},\omega )$ is the Hermitian part of the dielectric tensor and $\mathbf{v}_{g}
=\partial \omega _{0} /\partial \mathbf{k}_{0}$ is the group velocity of the wave.

Assuming the group velocities of the incident and scattered waves to be considerably larger than the electron thermal velocity,
we use cold plasma approximation. Within this model we write the expression for the linear dielectric tensor in the form \cite{17,18}
\begin{equation}
\varepsilon _{ij} (\omega )=\varepsilon _{1} (\omega )\delta _{ij}
-\varepsilon _{2} (\omega )b_{i} b_{j} +i\varepsilon _{3} (\omega )e_{ijl} b_{l} ,
\label{eq:11}
\end{equation}%
where $\mathbf{b}$ is the unit vector in the direction of the external magnetic field, $e_{ijl}$ is a fully antisymmetric unit
tensor, and
\begin{eqnarray}
&&\varepsilon _{1} (\omega )=1+\sum\limits_{a}\frac{\omega _{pa} }{\omega }
g_{a} (\omega ) ,\;\;\varepsilon _{2} (\omega )=\sum\limits_{a}\frac{\omega
_{pa} }{\omega } h_{a} (\omega ) ,  \label{eq:12} \\
&&\varepsilon _{3} (\omega )=\sum\limits_{a}\frac{\omega _{pa} }{\omega } l_{a} (\omega ) ,  \nonumber
\end{eqnarray}
\begin{eqnarray}
&&g_{a} (\omega )=\frac{\omega _{pa} (\omega +i\nu )}{\omega _{ca}^{2}
-\left( \omega +i\nu \right) ^{2} } ,\;\;l_{a} (\omega )=\frac{\omega _{ca}
\omega _{pa} }{\omega _{ca}^{2} -\left( \omega +i\nu \right) ^{2} } ,   \label{eq:13} \\
&&h_{a} (\omega )=\frac{\omega _{ca}^{2} \omega _{pa} }{\left( \omega +i\nu
\right) [ \omega _{ca}^{2} -\left( \omega +i\nu \right) ^{2}] } .  \nonumber
\end{eqnarray}
The summation in Eqs.~\eqref{eq:12} and \eqref{eq:13} is carried out over all plasma species $a$, $\omega _{pa}$ and
$\omega _{ca} =e_{a} B_{0} /m_{a} c$ are the plasma and cyclotron frequencies of particles of the kind $a$, and $\nu$ is the
effective frequency of electron-ion collisions.

For the vector $\mathbf{S}_{0}$ we obtain from Eqs.~\eqref{eq:10} and \eqref{eq:11}
\begin{equation}
\mathbf{S}_{0} =\frac{\mathbf{v}_{g} }{2\omega _{0} c} \frac{\partial }{%
\partial \omega _{0} } \Rea \left\{ \omega _{0}^{2} \left[
\varepsilon _{1} \left( \omega _{0} \right) -|\mathbf{b}\cdot \mathbf{e}%
|^{2} \varepsilon _{2} \left( \omega _{0} \right)
+i\varepsilon _{3} \left( \omega _{0} \right) \left( \mathbf{e}%
\cdot [\mathbf{e}^{*} \times \mathbf{b}\rbrack \right) \right] \right\} .
\label{eq:14}
\end{equation}

An expression for the tensor $S_{ipl}$ in the cold plasma approximation and in the absence of particle collisions was obtained
in \cite{18}. With allowance for the collisions, the expression for $S_{ipl}$ takes the form
\begin{eqnarray}
&&S_{ipl} (\mathbf{k},\omega ;\mathbf{k^{\prime} },\omega ^{\prime}
)=\sum\limits_{a}\frac{ie_{a} }{m_{a} } \frac{1}{\omega \omega ^{\prime}
\omega ^{\prime \prime} } \left[ \omega \Gamma _{il}^{(a)} (\omega )\Gamma
_{\alpha p}^{(a)} (\omega ^{\prime \prime} )k^{\prime} _{\alpha }
+\omega \Gamma _{ip}^{(a)} (\omega )\Gamma _{\alpha l}^{(a)} (\omega
^{\prime} )k^{\prime \prime} _{\alpha } -\omega ^{\prime} \Gamma
_{ij}^{(a)} (\omega )\Gamma _{pl}^{(a)} (\omega ^{\prime} )k^{\prime \prime} _{j}  \right. \label{eq:15} \\
&&\left. -\omega ^{\prime \prime} \Gamma _{ij}^{(a)} (\omega )\Gamma _{lp}^{(a)}
(\omega ^{\prime \prime} )k^{\prime} _{j} -\omega ^{\prime \prime} \Gamma
_{ip}^{(a)} (\omega ^{\prime \prime} )\Gamma _{sl}^{(a)} (\omega ^{\prime} )k_{s}
-\omega ^{\prime} \Gamma _{il}^{(a)} (\omega ^{\prime} )\Gamma _{sp}^{(a)} (\omega ^{\prime \prime} )k_{s} \right] ,  \nonumber
\end{eqnarray}
where $\Gamma _{ij}^{(a)} (\omega )=-g_{a} (\omega )\delta _{ij} +h_{a} (\omega )b_{i} b_{j} -il_{a} (\omega )e_{ijl} b_{l}$.
It should be noted that at $\omega ^{\prime \prime} =0$ ($\omega =\omega ^{\prime}$) the tensor S$_{ipl}$ has a singularity,
due to the adopted cold plasma model. Taking into account the thermal motion of the plasma particles the frequency change, in
the scattering process, is on the order of $\sim kv_{Te} =(v_{Te} /c)\omega _{0}$, where $v_{Te}$ is the electron thermal
velocity. We introduce the truncation parameter $T$, which is related to the frequency change $\omega ^{\prime \prime}$
by the relation $\omega ^{\prime \prime} =1/T$. Obviously at $\omega =\omega ^{\prime}$ we have $\omega _{0} T \sim c/v_{Te}$.

Let us consider the case when the test particle is at rest at a plane plasma-vacuum boundary. We consider the radiation
escaping from the plasma into the vacuum due to the scattering (or transformation) of the magnetized plasma waves on this
stationary particle. A more rigorous statement of the problem (boundary-value problem) requires that the generated surface
waves are also taken into account. However, it should be emphasized that their intensity decays exponentially with distance
from the boundary. Here we are interested only in the scattered bulk waves and the influence of the plasma boundary is
neglected.

Consider Eq.~\eqref{eq:9} for the intensity of the scattered radiation in the vacuum, where $\varepsilon _{ij} (\mathbf{k},\omega )
\to \delta _{ij} +i0$. For the tensor $T_{ij}$ in this limit we obtain
\begin{equation}
T_{ji} (\mathbf{k},\omega _{0} )-T_{ij}^{*} (\mathbf{k},\omega _{0} )=2\pi i%
\frac{\omega _{0}^{2} }{c^{2} } (\delta _{ij} -n_{i} n_{j} )\delta \left(
k^{2} -\frac{\omega _{0}^{2} }{c^{2} } \right) ,
\label{eq:16}
\end{equation}
where $\mathbf{n}=\mathbf{k}/k$ is the unit vector in the direction of the wave vector $\mathbf{k}$ of the scattered waves.
It thus follows from Eq.~\eqref{eq:16} that the scattering process occurs with no frequency change, while the wavelength differs
from that of the incident waves ($k/k_{0} =\omega _{0} /k_{0} c=\eta$) owing to the difference between the phase velocities of
the plasma waves and the speed of light in a vacuum.

Using Eqs.~\eqref{eq:9}, \eqref{eq:15}, and \eqref{eq:16} as well as the relation $\varepsilon (\mathbf{k},0)=1+(k\lambda _{D} )%
^{-2}$ for the static dielectric function, where $\lambda _{D}$ is the Debye screening length of the plasma, for the total intensity
of the scattered waves after lengthy but straightforward calculations we finally obtain
\begin{equation}
W_{s} =\int I(\mathbf{n},\mathbf{n}_{0} )d\Omega ,
\label{eq:17}
\end{equation}%
where $d\Omega =\sin \theta d\theta d\varphi$, $\cos \theta =\mathbf{n}\cdot \mathbf{n}_{0}$, $\theta$ is the scattering angle,
$\mathbf{n}_{0} =\mathbf{k}_{0} /k_{0}$ is the unit vector in the direction of the incident wave vector $\mathbf{k}_{0}$, and
$I(\mathbf{n},\mathbf{n}_{0} )$ is the angular distribution of the scattered radiation
\begin{equation}
I(\mathbf{n},\mathbf{n}_{0} )=\frac{I_{0} Z^{2} \left( \omega _{0} T\right)
^{2} \Im (\mathbf{n},\mathbf{n}_{0} )}{\left[ \eta ^{2} +1+\lambda ^{2}
/\lambda _{D}^{2} -2\eta (\mathbf{n}\cdot \mathbf{n}_{0} )\right] ^{2} } ,
\label{eq:18}
\end{equation}%
\begin{equation}
\Im (\mathbf{n},\mathbf{n}_{0} )=\sum\limits_{a;b}\mu _{a} \mu _{b} \Psi
^{(a)} (\mathbf{n},\mathbf{n}_{0} )\Psi ^{(b)} (\mathbf{n},\mathbf{n}_{0}
)\Phi _{ab} (\omega ,\mathbf{n},\mathbf{n}_{0} ) ,
\label{eq:19}
\end{equation}%
\begin{equation}
\Psi ^{(a)} (\mathbf{n},\mathbf{n}_{0} )=G_{a} (\eta \mathbf{n}-\mathbf{n}%
_{0} )^{2} +H_{a} [\eta (\mathbf{n}\cdot \mathbf{b})-(\mathbf{n}_{0} \cdot
\mathbf{b})\rbrack ^{2} ,
\label{eq:20}
\end{equation}%
\begin{eqnarray}
&&\Phi _{ab} (\omega ,\mathbf{n},\mathbf{n}_{0} )=g_{a} (\omega )g_{b}^{*}
(\omega )\left[ 1-|\mathbf{n}\cdot \mathbf{e}|^{2} \right] +l_{a} (\omega )l_{b}^{*} (\omega )
\left\{ \left| [\mathbf{e}\times \mathbf{b}\rbrack \right| ^{2}
-\left| \mathbf{n}\cdot [\mathbf{e}\times \mathbf{b}\rbrack \right| ^{2} \right\} \label{eq:21} \\
&& +g_{a} (\omega )h_{b}^{*} (\omega )(\mathbf{n}\cdot \mathbf{b})(%
\mathbf{n}\cdot \mathbf{e})(\mathbf{e}\cdot \mathbf{b})^{*}
+g_{b}^{*} (\omega )h_{a} (\omega )\left( \mathbf{n}\cdot \mathbf{b}\right)
\left( \mathbf{e}\cdot \mathbf{b}\right) \left( \mathbf{n}\cdot \mathbf{e} \right) ^{*} \nonumber \\
&&+i\left( \mathbf{n}\cdot \mathbf{e}\right) \left( \mathbf{n}%
\cdot [\mathbf{e}\times \mathbf{b}\rbrack ^{*} \right)
g_{a} (\omega )l_{b}^{*} (\omega )-i\left( \mathbf{n}\cdot \mathbf{e}%
\right) ^{*} \left( \mathbf{n}\cdot [\mathbf{e}\times \mathbf{b}\rbrack
\right) l_{a} (\omega )g_{b}^{*} (\omega ) \nonumber \\
&&+\left\{ h_{a} (\omega )h_{b}^{*} (\omega )[1-(\mathbf{n}\cdot \mathbf{b}%
)^{2} \rbrack -g_{a} (\omega )h_{b}^{*} (\omega )-h_{a} (\omega )g_{b}^{*}
(\omega )\right\} \left| \mathbf{e}\cdot \mathbf{b}\right| ^{2} \nonumber \\
&&+i\mathbf{e}\cdot [\mathbf{b}\times \mathbf{e}^{*} \rbrack \left[ l_{a} (\omega )g_{b}^{*}
(\omega )+g_{a} (\omega )l_{b}^{*} (\omega )\right]
+i\left( \mathbf{n}\cdot \mathbf{b}\right) \left[ \left( \mathbf{b}\cdot
\mathbf{e}\right) ^{*} \left( \mathbf{n}\cdot [\mathbf{e}\times \mathbf{b}%
\rbrack \right) l_{a} (\omega )h_{b}^{*} (\omega )\right. \nonumber \\
&&\left. -\left( \mathbf{b}\cdot \mathbf{e}\right) \left( \mathbf{n}\cdot \lbrack %
\mathbf{e}^{*} \times \mathbf{b}\rbrack \right) h_{a} (\omega )l_{b}^{*} (\omega )\right] . \nonumber
\end{eqnarray}
In Eqs.~\eqref{eq:18}-\eqref{eq:21} we have introduced the following notations: $\lambda =1/k_{0}$ is the wavelength of the incident wave,
$I_{0} =(c\left| \boldsymbol{\mathcal{E}}_{0} \right| ^{2} /2\pi )r_{0}^{2}$, $r_{0} =e^{2} /mc^{2}$ is the electron classical
radius, and $G_{a} =-ig_{a} (0)$, $H_{a} =ih_{a} (0)$, $\mu _{a} =me_{a} /m_{a} e$.

Below, in Secs.~\ref{sec:3} and \ref{sec:4}, in the case of scattering and transformations of the high-frequency plasma waves
we consider the
interaction of the incident wave only with the electron component of the plasma omitting the index a in expressions (19)-(21),
assuming that the quantities $g(\omega )$, $h(\omega )$, $l(\omega )$, $G$, and $H$ are related to the electrons. However, the
ion component of the plasma must be taken into account in the case of low-frequency incident waves when $\omega_{0} \sim \omega%
_{ci} ,\omega _{pi}$ (Secs.~\ref{sec:5} and \ref{sec:6}). Furthermore, we consider the general expressions (18)-(21) in some special cases,
assuming that the incident wave propagates perpendicular to the magnetic field $\mathbf{B}$ direction. We also assume that the
incident wave propagates perpendicular to the plasma-vacuum interface (i.e., we choose the magnetic field to be parallel to the
plasma boundary).

\section{Scattering of Ordinary Waves}
\label{sec:3}

We first consider the scattering of ordinary waves from a stationary charged particle. It is well known \cite{17} that an ordinary
wave is a linearly polarized transverse ($\mathbf{E}_{0} \perp \mathbf{k}_{0}$, where $\mathbf{E}_{0} =2\boldsymbol{\mathcal{E}}_{0}$)
electromagnetic wave propagating across a magnetic field. The polarization vector of this wave is parallel to the external magnetic
field, $\mathbf{E}_{0} \parallel \mathbf{B}$, while its frequency is related to the wave vector by the usual dispersion equation
for transverse electromagnetic waves propagating in a plasma, $\omega _{0}^{2} =\omega _{p}^{2} +k_{0}^{2} c^{2}$. The amplitude
of the magnetic field of the incident wave is determined by the relation $\mathbf{B}_{0} =(c/\omega _{0} )[\mathbf{k}_{0} \times
\mathbf{E}_{0} \rbrack$.

\begin{figure}[tbp]
\includegraphics[width=70mm]{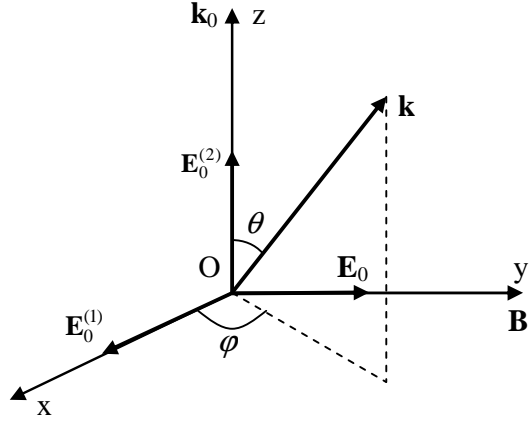}
\caption{Diagram illustrating the scattering of an ordinary wave from a
stationary charged particle located at the plane of a plasma-vacuum
interface. The wave is traveling perpendicular to the plasma surface toward
its boundary. The magnetic field is parallel to the interface and is
directed along the polarization vector of the incident wave.}
\label{fig:1}
\end{figure}

We introduce a spherical coordinate system with the polar $z$ axis in the direction of the vector $\mathbf{k}_{0}$ and the $y$
axis in the direction of the vectors $\mathbf{E}_{0}$ and $\mathbf{B}$ (Fig.~\ref{fig:1}). The angle $\varphi$ is determined from the
direction of the $x$ axis. Then, taking into account the dispersion law for the incident ordinary wave and neglecting the ion
component of the plasma, from Eq.~\eqref{eq:18} we obtain
\begin{equation}
I(\theta ,\varphi )=I_{0} \frac{Z^{2} \left( \omega _{0} T\right) ^{2} \Psi
^{2} (\theta ,\varphi )\Phi (\omega _{0} ,\theta ,\varphi )}{\left( \eta
^{2} +1+\lambda ^{2} /\lambda _{D}^{2} -2\eta \cos \theta \right) ^{2} } ,
\label{eq:22}
\end{equation}%
where $\eta ^{2} =1+\lambda ^{2} /\lambda _{p}^{2}$, $\lambda _{p} =c/\omega _{pe}$,
\begin{equation}
\Psi (\theta ,\varphi )=G\left( \eta ^{2} +1-2\eta \cos \theta \right)
+H\eta ^{2} \sin ^{2} \theta \sin ^{2} \varphi ,
\label{eq:23}
\end{equation}%
\begin{equation}
\Phi (\omega ,\theta ,\varphi )=\frac{\omega _{pe}^{2} }{\omega ^{2} +\nu
^{2} } \left( 1-\sin ^{2} \theta \sin ^{2} \varphi \right)
\label{eq:24}
\end{equation}%
and the angle $\theta$ varies in the range $0\leqslant \theta \leqslant \pi /2$. The collision frequency $\nu$ can be omitted
from Eq.~\eqref{eq:24}, since $\nu \ll \omega _{0}$ for any $k_{0}$.

The wave vector of the scattered wave is easily determined by equating to zero the argument of the delta function in Eq.~\eqref{eq:16},
$k=\omega _{0} /c>k_{0}$. This relation indicates that a long wavelength ordinary wave is transformed into short wavelength
electromagnetic radiation in a vacuum.

Let us briefly consider the results which follow from Eq.~\eqref{eq:22} in the absence of a magnetic field ($H=0$, $G=\omega _{pe}
/\nu$). In this case and for $\lambda <\lambda _{p}$ the radiation is concentrated mainly in the direction perpendicular to
the $zy$ plane, i.e., the scattered wave escapes into the vacuum almost parallel to the vacuum-plasma interface. For scattering
of a long waves ($\lambda >\lambda _{p}$) the radiation is uniformly distributed (i.e., it does not depend on the angle
$\theta$) in the $xz$ plane.

In the limit of wavelengths that are large compared to $\lambda _{p}$ the intensity of the scattered radiation does not depend
on wavelength, in accordance with Eqs.~\eqref{eq:22} and \eqref{eq:23}, and has the form
\begin{equation}
I(\theta ,\varphi )=I_{0} \frac{G^{2} Z^{2} \left( \omega _{pe} T\right)
^{2} }{\tau ^{4} } \left( 1-\sin ^{2} \theta \sin ^{2} \varphi \right)
\left( 1+\frac{H}{G} \sin ^{2} \theta \sin ^{2} \varphi \right) ^{2} ,
\label{eq:25}
\end{equation}
where $\tau =\lambda _{p} /\lambda _{D} \gg 1$. From Eq.~\eqref{eq:25} it is seen that in the absence of a magnetic field ($H=0$
) the scattering occurs just as from a point charge Ze (Thomson scattering) having an effective mass $m_{\text{eff} }
=Zm\tau ^{2} /(\omega _{pe} TG)$ \cite{16}. Thus, the term proportional to $H$ in Eq.~\eqref{eq:25} determines the scattering of long
waves due to plasma anisotropy.

\begin{figure}[tbp]
\includegraphics[width=60mm,angle=90]{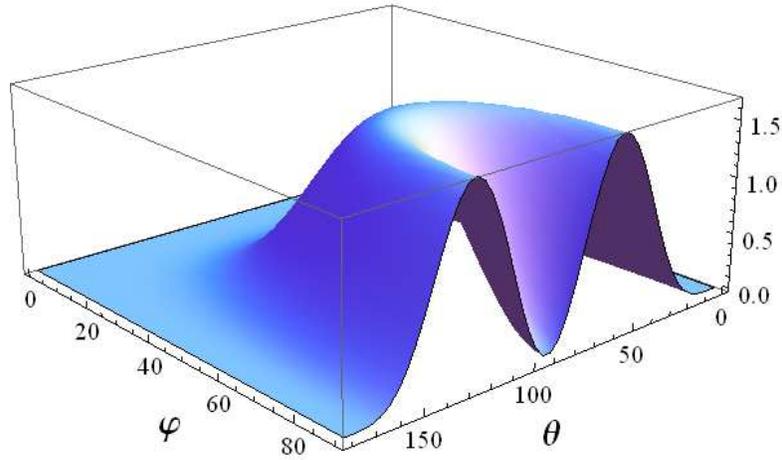}
\caption{(Color online) Angular distribution (normalized to $10^{-7}J_{0}$, where $J_{0} =I_{0}Z^{2} (\omega_{pe}T)^{2}$)
of the scattered ordinary wave in a long wavelength range ($\lambda = 4\lambda_{p}$). The calculations were done for $\tau
= 10^{2}$, $\nu /\omega_{pe}$ = 0.1, and $\omega _{ce}/\omega_{pe}$ = 3.}
\label{fig:2}
\end{figure}

It will be shown below that a sufficiently strong magnetic field ($\omega _{ce} \gg \nu$ or $H\gg G$) can significantly affect
the scattering pattern observed in the absence of an external magnetic field. The angular distribution of the intensity of
scattered radiation in this case has a maximum, the position of which is determined by the relation
\begin{equation}
(\mathbf{n}\cdot \mathbf{b})^{2} =\sin ^{2} \theta \sin ^{2} \varphi \simeq
\frac{2-\nu ^{2} /\omega _{ce}^{2} }{3} .
\label{eq:26}
\end{equation}%
From Eq.~\eqref{eq:26} it is seen that the maximum of the intensity exists only for sufficiently strong magnetic fields,
$\omega _{ce} >\nu /\sqrt{2}$. In the opposite case with $\omega _{ce} <\nu /\sqrt{2}$ the intensity decreases monotonically,
while for sufficiently small angle $\varphi$ ($\sin ^{2} \varphi <(2-\nu ^{2} /\omega _{ce}^{2} )/3$) but for
$\omega _{ce} >\nu /\sqrt{2}$ it increases monotonically with $\theta$. From Eqs.~\eqref{eq:25} and \eqref{eq:26} it follows that the maximum
of the intensity decreases slowly (by a factor of about 2.2) as the magnetic field increases from zero to the values of
$\omega _{ce} \gg \omega _{pe}$. The function $I(\theta ,\varphi )$ is shown in Fig.~\ref{fig:2} for the scattering of long waves
($\lambda =4\lambda _{p}$) as a function of $\theta$ and $\varphi$. It is seen that the scattered radiation is concentrated mainly
near a contour on $\theta$, $\varphi$ plane defined by Eq.~\eqref{eq:26}. We also note that these equations define two cones
$(\mathbf{n}\cdot \mathbf{b})^{2} =\text{const}$ with apices at the point $x = y = z = 0$ (see Fig.~\ref{fig:1}).

\begin{figure}[tbp]
\includegraphics[width=60mm,angle=90]{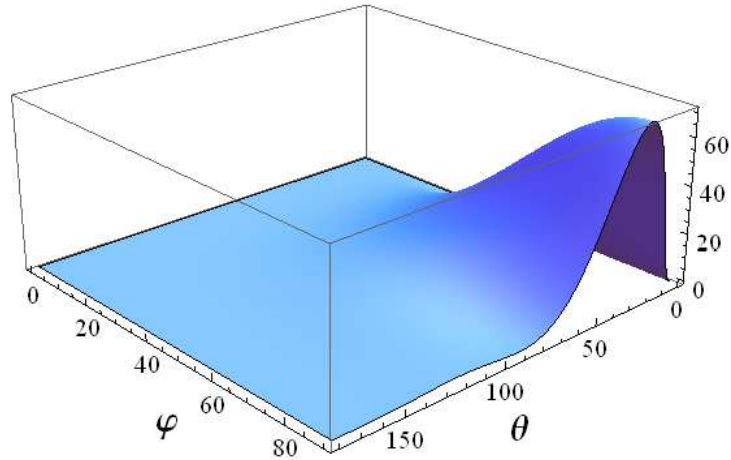}
\caption{(Color online) Angular distribution (normalized to $J_{0}$) of the scattered ordinary wave in a short wavelength
range ($\lambda = 10^{-3}\lambda_{p}$). The values of the other parameters are the same as in Fig.~\ref{fig:2}.}
\label{fig:3}
\end{figure}

With a decrease of the incident wave wavelength the intensity of the scattered radiation increases rapidly, approximately as
$\lambda ^{-4}$ (see the denominator of Eq.~\eqref{eq:22}), up to $\lambda \sim \lambda _{D}$. Here the intensity has a maximum,
the position of which is determined by Eq.~\eqref{eq:26}, in the wavelength range $\lambda _{D} <\lambda <\lambda _{p}$. It should
be noted that the features of the angular distribution of the scattered waves discussed above for $\lambda >\lambda _{p}$ are
retained in the case of small wavelengths.

In the limit of very short waves ($\lambda <\lambda _{D}$) the angular distribution $I(\theta ,\varphi )$ is changed significantly.
Under the condition $\omega _{ce} \gg \nu$, $\sin \varphi >\nu /\omega _{ce}$, for example, Eq.~\eqref{eq:22} takes the form
\begin{equation}
I(\theta ,\varphi )=I_{0} Z^{2} \left( \omega _{pe} T\right) ^{2} \left[
\frac{H\sin ^{2} \theta \sin ^{2} \varphi }{4\sin ^{2} (\theta /2)+\lambda
^{2} /\lambda _{D}^{2} } \right] ^{2}
\left( 1-\sin ^{2} \theta \sin ^{2} \varphi \right) .
\label{eq:27}
\end{equation}
The intensity maximum is shifted toward smaller $\theta$ in this case, while the position of that maximum is determined by
the equation
\begin{equation}
\theta _{\max } \simeq \left( \frac{2\lambda }{\lambda _{D} } \frac{1}{%
\sqrt{1+2\sin ^{2} \varphi } } \right) ^{1/2} .
\label{eq:28}
\end{equation}%
As follows from Eqs.~\eqref{eq:27} and \eqref{eq:28}, the maximum intensity increases rapidly with increasing $\varphi$. In
Fig.~\ref{fig:3} we demonstrate the angular distribution $I(\theta ,\varphi )$ for the scattering of the short waves
($\lambda =0.1\lambda _{D} \ll \lambda _{p}$). Thus, the scattering in this case occurs mainly in the direction of propagation
of the ordinary wave.

The total cross section for scattering from a stationary particle is obtained from Eq.~\eqref{eq:22} after integration over the
angles $\theta$ and $\varphi$, where for ordinary waves from Eqs.~\eqref{eq:12}-\eqref{eq:14} we obtain $v_{g} =c/\sqrt{1+\lambda ^{2} /
\lambda _{p}^{2} }$ and $S_{0} \simeq v_{g} /c$. Since the general expression for the cross section is cumbersome, below we
consider only some particular cases. In the case of scattering of very short waves ($\lambda \ll \lambda _{D}$) the cross
section is almost constant and is given by
\begin{equation}
\sigma (\lambda )\simeq \sigma _{T} \frac{Z^{2} }{2} \left( \omega _{pe}
T\right) ^{2} \left( a_{1} G^{2} +b_{1} GH+c_{1} H^{2} \right) ,
\label{eq:29}
\end{equation}%
where $\sigma _{T} =(8\pi /3)r_{0}^{2}$ is the Thomson cross section, $a_{1} =1$, $b_{1} =39/64$, $c_{1} =45/256$.

In the intermediate regime with $\lambda _{D} <\lambda <\lambda _{p}$ the cross section decreases as
$\sigma (\lambda )\simeq \sigma _{T} \sigma _{1} (\lambda _{p} /\lambda )^{4}$, where
\begin{equation}
\sigma _{1} =\frac{11Z^{2} }{20\tau ^{4} } \left( \omega _{pe} T\right)
^{2} \left( a_{2} G^{2} +b_{2} GH+c_{2} H^{2} \right)
\label{eq:30}
\end{equation}%
with $a_{2} =1$, $b_{2} =17/44$, $c_{2} =6/77$.

In the case of long wavelengths ($\lambda >\lambda _{p}$) the cross section increases linearly with the wavelength of the
incident wave, $\sigma (\lambda )\simeq \sigma _{T} \sigma _{2} \lambda /\lambda _{p}$, where
\begin{equation}
\sigma _{2} =\frac{Z^{2} }{2\tau ^{4} } \left( \omega _{pe} T\right) ^{2}
\left( a_{3} G^{2} +b_{3} GH+c_{3} H^{2} \right)
\label{eq:31}
\end{equation}%
with $a_{3} =1$, $b_{3} =2/5$, $c_{3} =3/35$. Such a behavior of the cross section is explained by the fact that the
incident and scattered waves have different group velocities, and for $\lambda >\lambda _{p}$ the energy flux in the incident
wave is $S\sim 1/\lambda$. Therefore, in the long wavelength range the total cross section does not coincide with the Thomson
cross section for scattering from a point particle with a mass $m_{\mathrm{eff}}$, as it occurs in the absence of a plasma boundary and
an external magnetic field \cite{18}.

Using Eqs.~\eqref{eq:30} and \eqref{eq:31}, the scattering cross section for ordinary waves at $\lambda >\lambda _{D}$ can be
represented in the approximate form
\begin{equation}
\sigma (\lambda )\simeq \sigma _{T} \left[ \sigma _{1} \left( \frac{\lambda
_{p} }{\lambda } \right) ^{4} +\sigma _{2} \frac{\lambda }{\lambda _{p} } \right] .
\label{eq:32}
\end{equation}%
From Eq.~\eqref{eq:32} it follows that at $\lambda_{\min}\simeq \lambda_{p}(4\sigma_{1}/\sigma_{2})^{1/5}$ the cross section has
a minimum, the value of which is given by $\sigma _{\min } \simeq 1.25\sigma _{T} (4\sigma_{1}\sigma_{2}^{4})^{1/5}$.

The dependence of the scattering cross section on the magnetic field can be traced from Eqs.~\eqref{eq:29}-\eqref{eq:32}. The cross section
decreases monotonically with increasing magnetic field. This behavior is especially pronounced for $\lambda >\lambda _{D}$
and is one order of magnitude over the range of variation of the magnetic field from zero to $\beta =\omega_{ce}/\omega_{pe} >1$.
The decrease of the cross section is due to the reduction of the transverse cyclotron motion of plasma electrons with the magnetic
field. In the limit of very strong magnetic fields ($\beta \gg 1$) the plasma behaves like a one-dimensional fluid, the motion of
which is confined to oscillations along magnetic field lines.

\section{Transformation of Extraordinary Waves}
\label{sec:4}

In this section we consider an extraordinary incident wave with a complex amplitude $\boldsymbol{\mathcal{E}}_{0}=(1/2)(\mathbf{E}%
_{0}^{(1)}-i\mathbf{E}_{0}^{(2)})$ (where $\mathbf{E}_{0}^{(1)}$ and $\mathbf{E}_{0}^{(2)}$ are the real amplitudes), propagating
across the magnetic field. It is well known \cite{17} that in general an extraordinary wave is elliptically polarized in the $xy$
plane (Fig.~\ref{fig:1}), i.e., $\boldsymbol{\mathcal{E}}_{0}\perp \mathbf{B}$. With no loss of generality, we choose the vectors
$\mathbf{E}_{0}^{(1)}$ and $\mathbf{E}_{0}^{(2)}$ such that $E_{0x}^{(1)}>0$ and $E_{0y}^{(1)}=E_{0z}^{(1)}=E_{0x}^{(2)}=E_{0y}^{(2)}
=0$ (Fig.~\ref{fig:1}). For this choice, the amplitude of the magnetic field of the incident wave is determined by the relation
$\boldsymbol{\mathcal{B}}_{0}=(c/2\omega_{0})[\mathbf{k}_{0}\times \mathbf{E}_{0}^{(1)}]$ and is directed along the external
magnetic field.

The relation between the components $E_{0x}^{(1)}$ and $E_{0z}^{(2)}$ is given by the equation \cite{17}
\begin{equation}
\frac{E_{0z}^{(2)}}{E_{0x}^{(1)}}=\frac{\varepsilon_{3}(\omega_{0})}{%
\varepsilon_{1}(\omega_{0})}\equiv P(\omega_{0}) ,
\label{eq:33}
\end{equation}
while the relation between the frequency and the wave vector is given by the dispersion equation for the extraordinary waves
\cite{17},
\begin{equation}
k_{0}^{2}=\frac{\omega_{0}^{2}}{c^{2}}\frac{2\varepsilon_{R}(\omega
_{0})\varepsilon_{L}(\omega_{0})}{\varepsilon_{R}(\omega_{0})+\varepsilon _{L}(\omega_{0})} ,
\label{eq:34}
\end{equation}
where $\varepsilon_{R}(\omega)=\varepsilon_{1}(\omega)-\varepsilon_{3}(\omega)$ and $\varepsilon_{L}(\omega)=\varepsilon_{1}%
(\omega)+\varepsilon _{3}(\omega)$.

The energy flux of the incident wave and the intensity of the scattered extraordinary waves are determined by the expressions
$\mathbf{S}=(c|\mathcal{E}_{0}|^{2}/2\pi)\mathbf{S}_{0}$, (14), and (18)-(21), respectively, where
\begin{equation}
\mathbf{S}_{0}=\frac{\mathbf{v}_{g}}{2\omega_{0}c}\frac\partial{\partial
\omega_{0}}\left[ \omega_{0}^{2}\varepsilon_{1}(\omega_{0})\frac {%
1+3P^{2}(\omega_{0})}{1+P^{2}(\omega_{0})}\right] ,
\label{eq:35}
\end{equation}
\begin{equation}
\Phi(\omega,\theta,\varphi)=q(\omega)\left[ 1-\sin^{2}\theta\cos^{2}%
\varphi+Q(\omega)\sin^{2}\theta\right] ,
\label{eq:36}
\end{equation}
\begin{equation}
q(\omega)=\frac{\omega_{pe}^{2}\left( \omega^{2}-\omega_{pe}^{2}\right) ^{2}%
}{\omega^{2}\left( \omega^{2}-\omega_{H}^{2}\right)
^{2}+\omega_{pe}^{4}\omega_{ce}^{2}},\quad Q(\omega)=\frac{%
\omega^{2}\omega_{ce}^{2}}{\left( \omega^{2}-\omega_{pe}^{2}\right) ^{2}} .
\label{eq:37}
\end{equation}

We investigate the expressions obtained for $I(\theta,\varphi)$ for high-frequency (electron) extraordinary waves. The ion
component of the plasma can be again neglected in this case. The two solutions of the dispersion Eq.~\eqref{eq:34} then have the
form \cite{17}
\begin{equation}
\omega_{0}^{(\pm)}(k_{0})=\left( \frac{f_{1}(k_{0})\pm\sqrt{%
f_{1}^{2}(k_{0})-4f_{2}(k_{0})}}2\right) ^{1/2} ,
\label{eq:38}
\end{equation}
where $f_{1}(k_{0})=\omega_{1}^{2}+\omega_{2}^{2}+k_{0}^{2}c^{2}$, $f_{2}(k_{0})=\omega_{1}^{2}\omega_{2}^{2}+\omega_{H}^%
{2}k_{0}^{2}c^{2}$, and $\omega_{H}^{2}=\omega_{ce}^{2}+\omega_{pe}^{2}$ is the upper hybrid frequency. The frequencies
$\omega_{1}$ and $\omega_{2}$ are the cutoff frequencies which are the solutions of the equations $\varepsilon_{L}
(-\omega_{2})=\varepsilon_{L}(\omega_{1})=0$ and $\varepsilon_{R}(-\omega_{1})=\varepsilon_{R}(\omega_{2})=0$, respectively,
and under the condition $\omega_{ce}\omega_{ci}\ll \omega_{pe}^{2}$ (which is fully justified for both laboratory and
astrophysical conditions) have the form \cite{17}
\begin{equation}
\omega_{2}=\frac{\omega_{ce}}2+\sqrt{\frac{\omega_{ce}^{2}}4+\omega_{pe}^{2}}%
,\quad\omega_{1}=\frac{\omega_{pe}^{2}}{\omega_{2}}<\omega_{2} .
\label{eq:39}
\end{equation}

In this section we consider only the scattering of the high-frequency mode $\omega _{0}^{(+)}$. We briefly recall (see also
Eq.~\eqref{eq:38}) that for this mode $\omega_{0}^{(+)}(k_{0})$ increases monotonically from $\omega_{0}^{(+)}(k_{0})=\omega_{2}$ at
$k_{0}\to 0$ to $\omega_{0}^{(+)}(k_{0})=k_{0}c$ at $k_{0}\to \infty$. Since $P(\omega_{0})>0$ (or $E_{0z}^{(2)}>0$) in
this frequency range the high-frequency wave has, in general, right-hand elliptic polarization in the
$xz$ plane (in the positive $y$ direction). In the case of long waves ($k_{0}\to 0$), $P(\omega_{2})=1$, the wave is almost
circularly polarized, whereas in the case of short waves ($k_{0}\to \infty$), $P(\omega_{0})\ll 1$, this mode consists of a
linearly polarized, transverse electromagnetic wave. In the latter case the only difference between an extraordinary and an
ordinary wave is that the polarization vector of an extraordinary wave is perpendicular to the external magnetic field.

The wave vector of the scattered wave is determined by the expression $k=\omega_{0}/c$. From Eq.~\eqref{eq:38} we conclude that
$\omega_{0}/c> k_{0}$ in the entire wavelength range of the incident wave. Thus, as in the case of an ordinary wave, the
transformation of extraordinary waves into electromagnetic radiation in a vacuum is accompanied by a decrease of the
wavelength.

Let us consider the angular distribution of the scattered waves in the limits of small and large $\lambda$. In the limit of
very short waves ($\lambda \ll \lambda_{D}$) the angular distribution has a maximum at the values of the small angle $\theta$
determined by Eq.~\eqref{eq:28}, in which $\sin \varphi$ is replaced by $\cos \varphi$. All the properties obtained for ordinary waves
in the range of $\lambda$ under consideration are retained in this case. In this limit the cross section is almost constant
and is determined by Eq.~\eqref{eq:29} in which the numerical coefficients are $a_{1}=1$, $b_{1}=61/64$, $c_{1}=381/1280$.

\begin{figure}[tbp]
\includegraphics[width=60mm,angle=90]{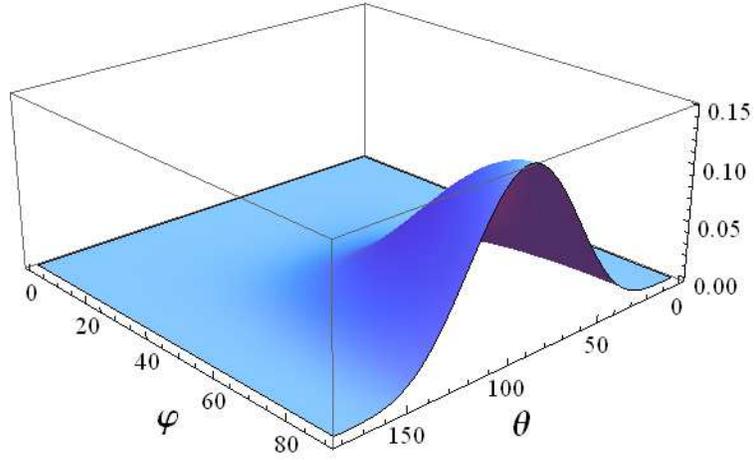}
\caption{Angular distribution (normalized to $J_{0}$) of a scattered extraordinary wave with a frequency $\omega _{0}^{(+)}$
in the intermediate wavelength range ($\lambda =5\lambda _{D}$). The values of the other parameters are the same as in
Fig.~\ref{fig:2}.}
\label{fig:4}
\end{figure}

In the intermediate wavelength range with $\lambda _{D}<\lambda \ll c\alpha /\omega _{H}$, where $\alpha $ is a number on the
order of unity, the intensity of scattered radiation decreases rapidly with increasing $\lambda $ as $\lambda ^{-4}$.
The angular distribution of the scattered waves is also changed. The intensity maximum is shifted toward larger angles $\theta $,
and for $\omega _{ce} \gg \nu $, $\sin \varphi >\nu /\omega _{ce}$, and $\cos ^{2}\varphi >2/3$ the position of that maximum
is determined by the expression $\sin ^{2}\theta \cos ^{2}\varphi \simeq 2/3$. However, $I(\theta ,\varphi )$ increases
monotonically with further increasing $\varphi $ ($\cos^{2} \varphi <2/3$) and reaches the maximum value at $\theta \simeq%
\varphi \simeq \pi /2$ (or $\varphi \simeq 3\pi /2$) (Fig.~\ref{fig:4}). In the same wavelength range the scattering cross section
has the form $\sigma (\lambda )\simeq \sigma _{T}\sigma _{1}(\lambda _{p}/\lambda )^{4}$, where $\sigma _{1}$ is determined
from Eq.~\eqref{eq:30} with coefficients $a_{2}=1$, $b_{2}=39/44$, and $c_{2}=18/77$.

In accordance with Eqs.~\eqref{eq:22}, \eqref{eq:23}, \eqref{eq:36}, and \eqref{eq:37} in the limit of the wavelengths larger
than $c/\omega_{ce}$, the intensity of the scattered radiation is
\begin{equation}
I(\theta,\varphi)=I_{0}Z^{2}(\omega_{pe}T)^{2}\frac{f_{0}^{8}(\beta)}{%
2\tau^{4}}\left( 1+\sin^{2}\theta\sin^{2}\varphi\right) \left( G+H\sin
^{2}\theta\sin^{2}\varphi\right) ^{2} ,
\label{eq:40}
\end{equation}
\begin{equation}
f_{0}(\beta)\equiv\frac{\omega_{2}}{\omega_{pe}}=\frac\beta2+\sqrt {1+\frac{ \beta^{2}}4} .
\label{eq:41}
\end{equation}
It is seen from (40) that the intensity of the scattered radiation increases monotonically with $(\mathbf{n} \cdot \mathbf{b})
=\sin\theta \sin \varphi$ and scattering occurs mainly in the direction of the external magnetic field.

In the limit of $\lambda\gg c/\omega_{ce}$, from Eq.~\eqref{eq:35} we obtain $S_{0}=(\lambda_{p}/\lambda) F_{1}(\beta)$, where
\begin{equation}
F_{1}(\beta)=\frac1{f_{0}^{2}(\beta)\sqrt{4+\beta^{2}}}\frac{f_{0}(\beta) (1+4\beta^{2})+%
5\beta /2}{f_{0}(\beta) (1+4\beta^{2})+3\beta} .
\label{eq:42}
\end{equation}
Then the cross section reads $\sigma(\lambda )\simeq \sigma_{T}\sigma_{2}\lambda/\lambda_{p}$, where $\sigma_{2}$ is
determined from Eq.~\eqref{eq:31} with coefficients $a_{3}=f_{0}^{8}(\beta)/F_{1}(\beta)$, $b_{3}=4a_{3}/5$, and
$c_{3}=9a_{3}/35$. From these expressions and from (40)-(42) it follows that at $\lambda \gg c/\omega_{ce}$ and $\beta\gg 1$
the angular distribution and the scattering cross section of the extraordinary waves are proportional to $\beta^{8}$ and $a_{3}=%
\beta^{11}$, respectively, and increase considerably with the magnetic field. At $\lambda\gg c/\omega_{ce}$ an
extraordinary wave has right-hand circular polarization in the $xz$ plane and at $\beta \gg 1$ its frequency is
$\omega_{2} \simeq \omega_{ce}$. Thus in this case a specific cyclotron resonance may occur which, however, differs from the
usual one so that the incident wave is polarized in the plane of incidence and propagates across the external magnetic field \cite{17}.

\section{Transformation of Waves with Intermediate Frequency}
\label{sec:5}

In this section we consider the transformation of the extraordinary waves with a frequency spectrum $\omega_{0}^{(-)}(k_{0})$
determined by Eq.~\eqref{eq:38}. The frequency $\omega_{0}^{(-)}(k_{0})$ increases monotonically (see Eq.~\eqref{eq:38}) from
$\omega_{0}^{(-)}=\omega_{1}$ as $k_{0}\to 0$ to $\omega_{0}^{(-)}=\omega_{H}$ at $k_{0}\to \infty$. Since the frequency of this
mode is high compared to the characteristic ionic frequencies (see, e.g., \cite{17}), we neglect here the contribution of plasma
ions both in the scattering current and in the wave dispersion. From Eq.~\eqref{eq:33} it follows that in this frequency range
$P(\omega_{0})<0$ (or $E_{0z}^{(2)}<0$), i.e., this mode, in general, has left-hand elliptical polarization in the $xz$ plane
(Fig.~\ref{fig:1}) and cannot resonate with plasma electrons. In the case of long waves ($k_{0}\to 0$), $P(\omega_{1})=-1$,
and the wave is almost circularly polarized, whereas in the case of short waves ($k_{0}\to \infty$), $P(\omega_{H})=-\infty$,
this mode consists of a longitudinal wave (upper-hybrid oscillations). In the latter case the upper-hybrid waves are transformed
into electromagnetic radiation in a vacuum.

From Eq.~\eqref{eq:38} for $\omega_{0}^{(-)}$ one concludes that $\omega_{0}/c\leqslant k_{0}$ at $\lambda\leqslant \lambda_{p}$
and $\omega_{0}/c >k_{0}$ at $\lambda> \lambda_{p}$. Thus, at $\lambda\leqslant \lambda_{p}$ and $\lambda>\lambda_{p}$ the scattering
of the mode $\omega_{0}^{(-)}$ is accompanied by an increase or a decrease of the wavelength, rspectively.

General expressions for the angular distribution of the scattered extraordinary waves have been obtained in Secs.~\ref{sec:3} and
\ref{sec:4} (Eqs.~\eqref{eq:22}, \eqref{eq:36}, and \eqref{eq:37}). In the range of very short wavelengths ($\lambda\ll \lambda_{D}$),
from these expressions we obtain the angular distribution of the transformation of upper-hybrid oscillations,
\begin{equation}
I(\theta,\varphi)=I_{0}Z^{2}(\omega_{pe}T)^{2}G^{2}\left( 1+\beta^{2}\right)
\left[ \beta^{2}+\sin^{2}\theta\left( \beta^{2}\sin^{2}\varphi+1\right) \right] .
\label{eq:43}
\end{equation}

The transformation cross section in this wavelength range it is obtained from Eqs.~\eqref{eq:35} and \eqref{eq:43}. After integration
of Eq.~\eqref{eq:43} with respect to the angles, one obtains $\sigma\simeq \sigma_{T}\sigma_{0}(\lambda_{p}/\lambda)^{3}$, where
\begin{equation}
\sigma_{0}=\frac{Z^{2}}3(\omega_{pe}T)^{2}G^{2}\sqrt{1+\beta^{2}}\left(
1+\frac1{2\beta^{2}}\right) .
\label{eq:44}
\end{equation}
From this expression for the cross section it is seen that, in contrast to the scattering (transformation) of high-frequency waves,
in which the cross section for $\lambda\ll \lambda_{D}$ is constant, in the case of intermediate upper-hybrid waves the cross section
increases essentially (as $\lambda^{-3}$) with decreasing the wavelength of the incident wave. This feature is due to the strong
reduction of the energy flux ($S_{0}\sim \lambda^{3}$) in the incident wave.

Consider now the opposite limiting case of the long wavelengths, $\lambda \gg\lambda_{p}$. We first note that for sufficiently strong
magnetic fields, $\omega_{ce}> \omega_{pe}/\sqrt{2}$ (or $\beta >1/\sqrt{2}$) the frequency of these waves at $\lambda =\lambda_{c}
=c/\sqrt{2\omega_{ce}^{2} -\omega_{pe}^{2}}$ coincides with the electron cyclotron frequency, $\omega_{0}^{(-)}=\omega_{ce}$. On the
other hand, $P(\omega) =-1$ at $\omega =\omega_{ce}$ and the incident wave is circularly polarized in the $xz$ plane (see Fig.~\ref{fig:1}).
Near the cyclotron frequency, $\omega_{0} \simeq \omega_{ce}$, the energy flux of the intermediate wave has the form
\begin{equation}
S_{0}\sim\frac{2\omega_{ce}\omega_{pe}^{4}\sqrt{2\omega_{ce}^{2}-\omega
_{pe}^{2}}}{\omega_{H}^{2}\left( \omega_{0}^{2}-\omega_{ce}^{2}\right) ^{2}}
\label{eq:45}
\end{equation}
and increases strongly due to cyclotron resonance. This resonance is stabilized taking into account the electron-ion collisions. Here
the energy flux can be very large but finite quantity. Thus, at $\lambda =\lambda_{c}$ the transformation cross section is vanishingly
small, $\sigma \simeq 0$.

In the limit of the long wavelengths, for the angular distribution from the general expressions \eqref{eq:22} and \eqref{eq:36} we
obtain
\begin{equation}
I(\theta,\varphi)=\frac{I_{0}Z^{2}(\omega_{pe}T)^{2}}{2\tau^{4}f_{0}^{8}(%
\beta)}\left( G+H\sin^{2}\theta\sin^{2}\varphi\right) ^{2}\left(
1+\sin^{2}\theta\sin^{2}\varphi\right) .
\label{eq:46}
\end{equation}
In this limit $S_{0} =(\lambda_{p}/\lambda)F_{2}(\beta)$, where
\begin{equation}
F_{2}(\beta)=f_{0}(\beta)\left\{ \frac{2f_{0}(\beta)}{\sqrt{4+\beta^{2}}}%
\left[ 1+\frac{\beta^{2}}{\left( 2-f_{0}^{2}(\beta)\right) ^{2}}\right]
-\frac12\right\} .
\label{eq:47}
\end{equation}

The cross section in the limit $\lambda \gg \lambda_{p}$ is determined from the expression $\sigma (\lambda)\simeq \sigma_{T}\sigma_{2}%
\lambda/\lambda_{p}$, where $\sigma_{2}$ is given by Eq.~\eqref{eq:31} with $a_{3}=1/f_{0}^{8}(\beta)F_{2}(\beta)$, $b_{3}=4a_{3}/5$, and
$c_{3}=9a_{3}/35 $. A comparison of these expressions for the angular distribution and the cross section with the similar expressions
obtained in the case of the transformation of a high-frequency extraordinary wave shows that in the former case a strong external magnetic
field can strongly suppress the transformation of an intermediate wave, the intensity of which decreases as $\beta^{-8}$ with increasing
of the external magnetic field (see Eq.~\eqref{eq:46}).

From Eqs.~\eqref{eq:1} and \eqref{eq:4} it is seen that the intensity of the transformation of an intermediate wave increases monotonically
with $\theta $ and takes a maximum value at $\theta =\pi /2$ and $\varphi =\pi /2$ (or $\varphi =3\pi /2$). Therefore, the radiation mainly
escapes from the plasma parallel to its boundary in the direction of the external magnetic field (Fig.~\ref{fig:1}).

\begin{figure}[tbp]
\includegraphics[width=60mm,angle=270]{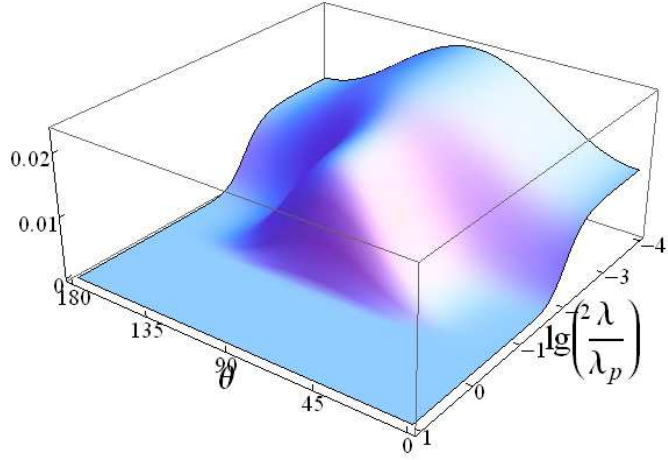}
\caption{Dependence of the intensity $I(\theta ,\varphi )$ (normalized to $J_{0}$) for a transformed intermediate wave on the
wavelength and the angle $\theta $ for $\varphi =\pi /2$. The values of the other parameters are the same as in Fig.~\ref{fig:2}.}
\label{fig:5}
\end{figure}

The intensity $I(\theta,\varphi)$ of the transformation of the intermediate wave as a function of the wavelength and the angle
$\theta$ is shown in Fig.~\ref{fig:5}. From this figure it is seen that the intensity has a maximum in the short-wavelength range,
whereas the intensity of the scattering (transformation) of the high-frequency waves decreases monotonically with wavelength of
the incident wave.

At the end of this section we note that the restriction $v_{g} >v_{Te}$ on the group velocity (see Sec.~\ref{sec:2}) leads to a
limitation $\lambda <(c/v_{Te}) \lambda_{p}/f_{0}^{2}(\beta) \sqrt{4+\beta^{2}}$ of the wavelength of the incident wave.

\section{Scattering of Low-Frequency Waves}
\label{sec:6}

In this section we consider the scattering (transformation) of low-frequency magnetosonic and lower-hybrid plasma waves,
the frequencies of which are much lower than the characteristic electron frequencies ($\omega_{ce}$ and $\omega_{pe}$)
and are comparable in order of magnitude with the ion-cyclotron and Langmuir frequencies $\omega_{ci}$ and $\omega_{pi}$,
respectively. In this low-frequency limit one must take into account the dynamics of the plasma ions and their partial
contributions to the dispersion equation and the scattering current.

From the general Eq.~\eqref{eq:34} we obtain an expression for the frequency of the low-frequency waves (see also \cite{17}),
\begin{equation}
\omega_{0}^{2}(k_{0})=\omega_{LH}^{2}\frac{k_{0}^{2}u_{A}^{2}}{k_{0}^{2}u_{A}^{2}+\omega_{LH}^{2}} ,
\label{eq:48}
\end{equation}
where $\omega_{LH}^{2}=\omega_{ce}\omega_{ci}\omega_{pe}^{2}/\omega_{H}^{2}$ is the lower hybrid frequency, $u_{A}=V_{A}%
/\sqrt{1+V_{A}^{2}/c^{2}}$, and $V_{A}$ is the Alfv\'{e}n velocity.

From Eq.~\eqref{eq:48} it follows that $P(\omega_{0})>0$ (or $E_{0z}^{(2)}>0$), i.e., this mode in general has right-hand elliptical
polarization in the $xz$ plane (Fig.~\ref{fig:1}) and can resonate with plasma ions. In the case of the long magnetosonic
waves ($k_{0}\to 0$) we obtain $P(0)=0$ and the wave has transverse polarization, while in the case of short lower-hybrid
waves ($k_{0}\to \infty$), $P(\omega_{LH})\to \infty$ and this mode consists of a longitudinal wave. In the latter case we
have the transformation of the lower-hybrid waves into electromagnetic radiation in a vacuum.

From the expression for $\omega_{0}$ it follows that $\omega_{0}/c <k_{0}$ for any $\lambda$. The transformation of the
low-frequency mode is therefore accompanied by an increase in wavelength. Let us make some estimates. In astrophysical
conditions for a density $\rho =10^{6}$ g/cm$^{3}$ and a magnetic field $B= 10^{9}$ kG we obtain that the transformation of
the magnetosonic wave in a vacuum generates radiation with a wavelength exceeding that of the incident wave by two orders of
magnitude, $k_{0}c/\omega_{0}=c/u_{A} \simeq 106$.

Consider now Eqs.~\eqref{eq:18}-\eqref{eq:21} for the intensity of the transformation of the low-frequency waves. Taking into account the
dynamics of the plasma ions Eqs.~\eqref{eq:19}-\eqref{eq:21} become
\begin{eqnarray}
&&\Im(\omega,\theta,\varphi) =\Psi^{(e)2}(\theta,\varphi)\Phi_{ee}(\omega,\theta,\varphi)+2\mu\Psi^{(e)}(%
\theta,\varphi)\Psi^{(i)}(\theta,\varphi)\Phi_{ei}(\omega,\theta,\varphi)+ \label{eq:49} \\
&&+\mu^{2}\Psi^{(i)2}(\theta,\varphi)\Phi_{ii}(\omega,\theta,\varphi ) , \nonumber
\end{eqnarray}
\begin{equation}
\Psi^{(a)}(\theta,\varphi)=G_{a}(\eta^{2}+1-2\eta\cos\theta) + \eta^{2}H_{a}\sin^{2}\theta\sin^{2}\varphi ,
\label{eq:50}
\end{equation}
\begin{equation}
\Phi_{ab}(\omega,\theta,\varphi)=q_{ab}(\omega)\left[ 1-\sin^{2}\theta
\cos^{2}\varphi+Q_{ab}(\omega)\sin^{2}\theta\right] ,
\label{eq:51}
\end{equation}
\begin{equation}
q_{ab}(\omega)=\frac{\left[ g_{a}(\omega)-l_{a}(\omega)P(\omega)\right] %
\left[ g_{b}(\omega)-l_{b}(\omega)P(\omega)\right] }{1+P^{2}(\omega)} ,
\label{eq:52}
\end{equation}
\begin{equation}
Q_{ab}(\omega)=\frac{\left[ l_{a}(\omega)-g_{a}(\omega)P(\omega)\right] %
\left[ l_{b}(\omega)-g_{b}(\omega)P(\omega)\right] }{\left[
g_{a}(\omega)-l_{a}(\omega)P(\omega)\right] \left[ g_{b}(\omega)-l_{b}(%
\omega)P(\omega)\right] } .
\label{eq:53}
\end{equation}
In Eqs.~\eqref{eq:50}-\eqref{eq:53} the indices $a$ and $b$ take the values $e$ or $i$ to denote the contributions of plasma
electrons and ions, $\mu =\mu_{i}=Z_{i}m/m_{i}\ll 1$, where $Z_{i}$, $m_{i}$, and $m$ are the charge number and mass of an
ion and the electron mass, respectively, and the quantities $G_{a}$, $H_{a}$, $g_{a}(\omega)$, $l_{a}(\omega)$, and $P(\omega)$
are determined by Eqs.~\eqref{eq:13} and \eqref{eq:33}, respectively.

In the limit of short wavelengths ($\lambda\ll \lambda_{D}$), from Eqs.~\eqref{eq:49}-\eqref{eq:53} we obtain the angular
distribution of the transformation of lower-hybrid oscillations,
\begin{equation}
I(\theta,\varphi)=I_{0}Z^{2}(\omega_{pe}T)^{2}\frac\mu{1+\beta^{2}}\left[
G_{1}^{2}\left( 1-\sin^{2}\theta\cos^{2}\varphi\right) +\frac\mu{1+\beta ^{2}%
}G_{2}^{2}\sin^{2}\theta\right] ,
\label{eq:54}
\end{equation}
where
\begin{eqnarray}
G_{1}=G_{e}+\mu^{3/2}\left( 1+\beta^{2}\right) G_{i} , \label{eq:55} \\
G_{2}=G_{e}-\mu^{1/2}\left( 1+\beta^{2}\right) G_{i} . \nonumber
\end{eqnarray}
For small external magnetic fields ($\beta\ll 1$) the contribution of the ions in Eqs.~\eqref{eq:54} and \eqref{eq:55} is
negligible. For $\beta\gg 1$, however, the transformation occurs mainly due to the ionic current.

From Eq.~\eqref{eq:54} we derive the cross section for the transformation of lower-hybrid waves by integrating the latter over angles. The
result is given by the expression $\sigma(\lambda) \simeq \sigma_{T}\sigma_{0} (\lambda_{p}/\lambda)^{3}$, where
\begin{equation}
\sigma_{0}=\frac{Z^{2}}6(\omega_{pe}T)^{2}\frac{\beta\sqrt{\mu}}{\sqrt {%
1+\beta^{2}}}\left( G_{1}^{2}+\frac\mu{1+\beta^{2}}G_{2}^{2}\right) .
\label{eq:56}
\end{equation}

From Eqs.~\eqref{eq:35} and \eqref{eq:48} it follows that for the wavelengths $\lambda \simeq \lambda_{c}=u_{A}/\omega_{ci}$ the
frequency is close to the ion-cyclotron frequency $\omega_{0} \simeq \omega_{ci}$. At these frequencies we obtain $P(\omega_{ci})=1$.
Therefore, near $\omega_{ci}$ the wave has right-hand circular polarization and, as noted above, it can resonate with plasma ions.
Ion-cyclotron resonance occurs in this case and the energy flux of the incident wave increases sharply as
\begin{equation}
S_{0}\sim\frac{u_{A}}c\frac{2\omega_{ci}^{2}\omega_{pi}^{2}}{\left(
\omega_{0}^{2}-\omega_{ci}^{2}\right) ^{2}} .
\label{eq:57}
\end{equation}
The cross section of the process thus approach to zero at $\lambda \simeq \lambda_{c}$.

In the limit of long magnetosonic waves ($\lambda \gg \lambda_{p}$, $\lambda_{c}$), from Eqs.~\eqref{eq:18} and \eqref{eq:49}-\eqref{eq:53}
we obtain
\begin{equation}
I(\theta,\varphi)=I_{0}Z^{2}(\omega_{pe}T)^{2}\frac\mu{\tau^{4}}\left( \frac{%
\lambda_{p}}\lambda\right) ^{6}\sin^{2}\theta\left[ G_{0}\left(
\eta_{A}^{2}+1-2\eta_{A}\cos\theta\right)
+H_{0}\eta_{A}^{2}\sin^{2}\theta\sin^{2}\varphi\right] ,
\label{eq:58}
\end{equation}
where $H_{0}=H_{e}-H_{i}\sqrt{\mu}$, $G_{0}=G_{e}-G_{i}\sqrt{\mu}$, and $\eta_{A}=u_{A}/c$. It should be noted that in the second
term in Eq.~\eqref{eq:58} the contribution of ions can be neglected for any values of the magnetic field ($H_{e} \gg H_{i}\sqrt{\mu}$),
whereas in the first term the contribution of ions can be neglected only for weak external magnetic fields.

From Eq.~\eqref{eq:58} we find the cross section for the transformation of magnetosonic waves. After evaluation of the integrals over
angles, we find $\sigma \simeq \sigma_{T}\sigma_{1}(\lambda_{p}/\lambda)^{6}$, where
\begin{equation}
\sigma_{1}=Z^{2}(\omega_{pe}T)^{2}\frac{\beta\mu^{3/2}}{2\tau^{4}}\left[
G_{0}\left( \eta_{A}^{2}+1-\frac34\eta_{A}\right) +\frac25H_{0}\eta_{A}^{2}\right] .
\label{eq:59}
\end{equation}

From Eqs.~\eqref{eq:54} and \eqref{eq:58} it is seen that the intensity of the transformation of the magnetosonic waves increases
monotonically with angles $\theta $ and $\varphi $, approaching the maximum value at $\theta =\pi /2$ and $\varphi =\pi /2$ (or
$\varphi = 3\pi /2$). Consequently, as in the case of an intermediate wave, radiation escapes from the plasma mainly parallel to its
boundary in the direction of the external magnetic field.

\begin{figure}[tbp]
\includegraphics[width=60mm,angle=90]{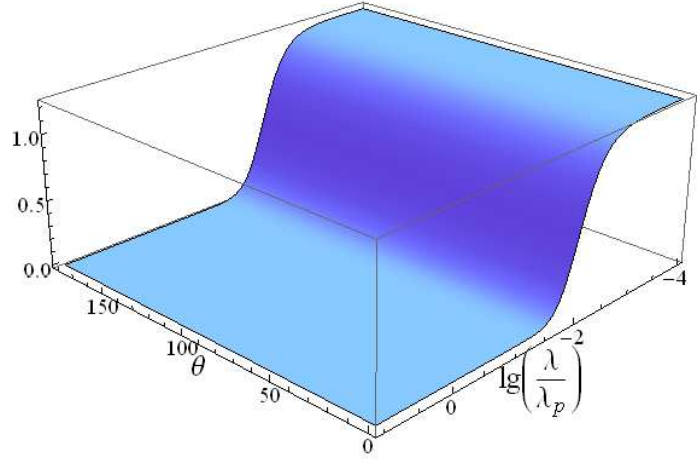}
\caption{Dependence of $I(\theta ,\varphi )$ (normalized to $10^{-10} J_{0}$) for a transformed low-frequency wave on the
wavelength and the angle $\theta $ for $\varphi =\pi /2$ and $\mu =10^{-5}$. The values of the other parameters are the
same as in Fig.~\ref{fig:2}.}
\label{fig:6}
\end{figure}

In Fig.~\ref{fig:6} we demonstrate the dependence of the intensity of transformation of a low-frequency wave as a function
of wavelength and the angle $\theta$. From (54), (58), and Fig.~\ref{fig:6} it is seen that the efficiency of transformation
of lower-hybrid waves far exceeds the efficiency of transformation of magnetosonic waves. However, the intensity of the
emission produced by the transformation of magnetosonic waves can be essential if we take into account that it is proportional
to the (large) quantities $Z^{2}$ and $T^{2}$. We note that the estimate of the cutoff parameter $T$ depends on the specific
model of magnetized plasma.

\section{Discussion and Conclusion}
\label{sec:7}

In this paper, we have presented a detailed investigation of the scattering
and transformation of the plasma waves on heavy charged particle in
magnetized plasma. The basic idea of this paper is that the scattering
(transformation) occurs due to the nonlinear interaction of the incident
wave with the polarization cloud surrounding the particle. In the course of
this study we have derived some analytical results for the angular
distribution and the cross section of the scattered (transformed) radiation
and we have shown that the problem is reduced to the determination of the
nonlinear (three index) dielectric tensor of magnetized plasma.

After introduction to the general theory in Sec.~\ref{sec:2}, we have
studied some particular cases of the scattering and transformation processes
assuming that the incident wave propagates in the direction transverse to
the external magnetic field. The angular distribution and the cross section
for the scattering and transformation of high-frequency ordinary and
extraordinary waves and low-frequency upper-hybrid, low-hybrid, and
magnetosonic waves have been investigated within a cold plasma model which
is valid when the group velocities of the incident and scattered waves
exceed the thermal velocities of the plasma particles. A number of limiting
and asymptotic regimes of short and long wavelengths have been studied. The
theoretical expressions for the angular distribution of the scattered waves
derived in this paper lead to a detailed presentation of a collection of
data through figures.

We expect our theoretical model to be useful in experimental
investigations of the wave scattering by plasma as well as in some
astrophysical applications. Going beyond the presented model calculations
which are based on the cold plasma approximation we can envisage a number of
avenues. One of the improvements of our model will be to include the thermal
effects which are particularly important in the case of dusty plasmas \cite{15}.
Furthermore, the theoretical model developed here although is strong but is
not adopted for immediate astrophysical applications. For this purpose it is
required (i) fully relativistic fluid calculations with appropriate equation
of states and transport coefficients of a strongly magnetized and dense
(degenerated) plasma (see, e.g., \cite{19}). (ii) Short range quantum effects
which appears due to the tunneling of electrons and positrons through the
Bohm quantum potential barrier \cite{20}. A study of these and other aspects will
be reported elsewhere.

\begin{acknowledgments}
This work has been supported by the Armenian Ministry of Higher Education
and Science under Grant No. 87.
\end{acknowledgments}

\end{document}